\documentclass[usenatbib]{mnras}
\usepackage{graphicx}
\usepackage{txfonts}
\usepackage{threeparttable}

\usepackage{aas_macros}
\title[Super-Eddington growth of the first black holes]{Super-Eddington growth of the first black holes}
\author[Pezzulli et al.]{Edwige Pezzulli$^{1,2,3}$\thanks{E-mail:
edwige.pezzulli@oa-roma.inaf.it}, Rosa Valiante$^{2}$ and Raffaella Schneider$^{2}$ \\
$^{1}$Dipartimento di Fisica, Universit{\'a} di Roma  ``La Sapienza'', P.le Aldo Moro 2, 00185, Roma, Italy \\
$^{2}$INAF/Osservatorio Astronomico di Roma, Via di Frascati 33, 00040 Monte Porzio Catone, Italy\\
$^{3}$INFN, Sezione di Roma I, P.le Aldo Moro 2, 00185 Roma, Italy }
\begin{document}

\date{1 March 2016}

\pagerange{\pageref{firstpage}--\pageref{lastpage}} \pubyear{2016}

\maketitle
\label{firstpage}

\begin{abstract}
The assembly of the first super massive black holes (SMBHs) at $z \gtrsim 6$  is still a subject
of intense debate. If black holes (BHs) grow at their Eddington rate, they
must start from $\gtrsim 10^4 \, M_\odot$ seeds formed by the direct collapse of gas.
Here we explore
the alternative scenario where $\sim 100 \, M_\odot$ BH remnants of the first stars grow at super-Eddington rate
via radiatively inefficient slim accretion disks.
We use an improved version of the cosmological, data-constrained semi-analytic model
\textsc{GAMETE/QSOdust}, where we follow the evolution of nuclear BHs and
gas cooling, disk and bulge formation of their host galaxies.
Adopting SDSS J1148+5251 (J1148)
at $z = 6.4$ as a prototype of luminous $z \gtrsim 6$ quasars, 
we find that $\sim$ 80\% of its SMBH mass 
is grown by super-Eddington accretion, which can be sustained down to $z \sim 10$ in dense, gas-rich
environments. The average BH mass at $z \sim 20$ is $M_{\rm BH} \gtrsim 10^4 \,M_\odot$,
comparable to that of direct collapse BHs.
At $z = 6.4$ the AGN-driven mass outflow rate is consistent with the observations and the BH-to-bulge mass ratio is
compatible with the local scaling relation.
However, the stellar mass in the central 2.5~kpc is closer
to the value inferred from CO observations.
Finally, $\sim 20 \%$ of J1148 progenitors at $z=7.1$ have BH luminosities and masses comparable to
ULAS J1120+0641, suggesting that this quasar may be one of the progenitors of J1148.

\end{abstract}

\begin{keywords}

quasars: supermassive black holes - galaxies: evolution - galaxies: high-redshift - galaxies: active - black hole physics - accretion, accretion discs

\end{keywords}

\section{Introduction}

High-redshift quasars are among the most luminous sources in the distant Universe. 
Their large luminosities ($L \sim 10^{47}$ erg/s) suggest that the powering mechanism of the strong 
radiative emission is the accretion of gas onto a Super Massive Black Hole (SMBH), with a mass 
$M_{\rm BH} \gtrsim 10^9 M_{\odot}$ settled in the center of the host galaxy \citep[e.g.][]{Fan01, Fan03, Willott07}.
This phenomenon, in fact, can convert up to $30\%$ of the energy in radiation, explaining the nature of 
this powerful emission \citep{Shakura73}.

The most distant quasars are observed up to redshifts $z \sim 7$ \citep{Mortlock11}, corresponding to a 
Universe younger than 1 Gyr old. How these SMBHs form and grow in such a short time is still an open 
question.

In the hierarchical scenario of structure formation, SMBHs are expected to grow via mergers with other
BHs and gas accretion, starting from a less massive BH, generally referred to as BH seed. 
Hence, the formation and accretion history of SMBHs depend on the initial mass of BH seeds and on their
formation epoch.
The nature of the first BH seeds is still uncertain and different formation mechanisms have been proposed in the literature \citep[see e.g.][and references therein]{Volonteri10}:
\begin{enumerate}
\item primordial black holes, with masses ranging from the Planck mass up to $10^5 M_{\odot}$ could have 
formed in the early Universe, well before galaxy formation \citep{Khoplov05}; 

\item remnants of the first generation of metal-poor stars, the so-called population III (Pop III) stars (see e.g. Bromm 2013 for a review), that can form to black holes of $\sim 100 \, M_{\odot}$, at $z \sim 20$ \citep{Madau01,Abel02, Bromm02, Oshea07, Turk09, Tanaka09, Greif12,Valiante16};

\item gas-dynamical processes in massive environment can lead to the direct collapse of gas into a massive BH of [$10^4 - 10^6$] M$_\odot$ \citep{Bromm03, Koushiappas04, Begelman06, Lodato06, Ferrara14, Pacucci15,Valiante16};

\item stellar-dynamical processes allow BHs to form in nuclear clusters of second generation stars with masses ~$\sim [10^2 - 10^5] M_{\odot}$ \citep{Devecchi09, Devecchi10, Devecchi12};

\item gas-driven core-collapse of dense stellar clusters due to the rapid infall of gas with a mass comparable to that of the stellar cluster can lead to the formation of BHs of $\sim 10^3 M_\odot$ or larger 
\citep{Davies11, Lupi14}.
\end{enumerate}

In order to grow up to billion solar masses at $z \sim 6$, seed BHs must accrete gas at the Eddington rate almost uninterruptedly for several hundreds Myr, even if they start as ``heavy seeds" of $[10^5 - 10^6] \, M_{\odot}$. Alternatively, short episodes of super-Eddington accretion have been suggested as a viable way to allow the efficient growth of SMBHs, especially if these start
from ``light seeds'' of $\sim 100 \, M_\odot$ \citep{Haiman04,Yoo04, Shapiro05, Volonteri05, Volonteri06, Pelupessy07, Tanaka09, Madau14, Volonteri15}.
In a recent numerical study, \citet{Lupi15} show that, if a large reservoir of dense cold gas is available, a $M_{\rm BH} \sim 10^5 M_{\sun}$ can grow  in a $\sim \rm Myr$ timescale 
starting from a seed mass of $\sim 20-100 \, M_{\odot}$, under the assumption of a slim accretion disk solution (Abramowicz et al. 1988). 
The slim disk solution, that we better describe in Section \ref{Sec:bh growth}, represents an advective, optically thick 
flows that generalise the standard Shakura $\&$ Sunyaev solution \citep{Shakura73}. In this model, the radiative efficiencies, that depend on the accretion rate,  are low: the radiation is trapped and advected inward by the accretion flow (see however the recent simulations by \citealt{Sadowski16}).
In this scenario, the outflow has a negligible effect and the BH can accrete up to $80\% - 100\%$ of the gas mass available \citep{Pacucci15}.

Indeed, there is observational evidence of mildly super-critical 
accretion \citep{Kelly13, Page14} in quasars at redshift up to $\sim$ 7. 
In addition, recent numerical simulations aimed to study super-Eddington accretion onto a rapidly rotating BH \citep{McKinney14} and the energy, momentum and mass outflow rates from radiatively inefficient accretion discs \citep{Sadowski13} predict Eddington ratios $\eta_{\rm Edd} = L/L_{\rm Edd}$ up to 10, where $L_{\rm Edd}$ is the Eddington luminosity, defined as:

\begin{equation}
L_{\rm Edd} = \frac{4\pi G M_{\rm BH} m_p c}{\sigma_T} \simeq 3.3 \times 10^{10} \left( \frac{M_{\rm BH}}{10^6 M_{\odot}} \right) L_{\odot}
\end{equation}
\noindent
with $M_{\rm BH}$ the central BH mass, $m_p$ the proton mass, $c$ the speed of light and $\sigma_T$ the Thomson scattering cross section.
Such a high ratio has been also invoked to explain the nature of ultraluminous X-ray sources \citep[e.g.][]{Middleton13}. 

In this paper, we investigate the role of super-Eddington accretion in the formation of the first 
SMBHs at redshift $z\sim 6$, with the aim to understand what are the environments where it can occur and discuss the implications for the coevolution of the SMBHs and their host galaxies at high redshifts. We base our analysis on the data-constrained semi-analytical model \textsc{GAMETE/QSOdust} that allows to simulate a large number of hierarchical histories of a quasar host dark matter halo, following the star formation history, chemical evolution and nuclear black hole growth in all its progenitor galaxies. The model has been first successfully used to investigate the properties of the $z = 6.4$ quasar SDSS J1148+5251 by \citet{Valiante11,Valiante12}, applied to a sample of quasars at $5 < z < 6.4$ by \citet{Valiante14} and more recently used to investigate the relative importance of light and heavy
seeds in the early growth of high-z SMBHs under the assumption of Eddington-limited accretion \citep{Valiante16}. Here we present an improved version
of the model, that has been modified to follow gas cooling, disk and bulge formation, and BH gas accretion in all the progenitor systems of a $z = 6.4$ quasar, using
SDSS J1148+5251 (hereafter J1148) as a prototype for the general class of luminous high-redshift quasars.

The paper is organized as follows. 
In Section \ref{sec:model} we introduce the model, describing assumptions and physical 
prescriptions. In Section \ref{sec:results} we present the results. Finally, a discussion and the main conclusions are given in Section \ref{sec:discussion}.


\begin{table*}
\begin{center}
\caption{Observed and inferred properties of the quasar SDSS J1148+5251. The black hole mass, $M_{\rm BH}$, is estimated from the $\rm Mg_{\rm II}$ doublet and the $\lambda = 3000 \, \AA$ continuum \citep{Derosa11}. The mass of molecular gas, $M_{\rm H_2}$, and the dynamical mass, $M_{\rm dyn}\sin^2 i$, have been estimated from CO observations 
(see \citealt{Valiante14} for more details). The star formation rate, SFR, has been computed from the far-infrared (FIR) luminosity using the Kennicutt relation {(see Section \ref{sec:results} fore further details)}. 
The value of $L_{\rm FIR}$ and $M_{\rm dust}$ have been computed by \citet{Valiante11,Valiante14}. The bolometric luminosity $L_{\rm bol}$ is estimated from the observed flux at $1450 \, {\AA}$ \citep{Fan03}
using the bolometric correction by \citet{Richards06}.}\label{Tab1}
\scalebox{0.86}{
\begin{tabular}{|c|c|c|c|c|c|c|c|c|}\hline 
{$z$} & {$M_{\rm BH} \, [10^{9}M_{\odot}$]} &{$M_{\rm H2} \, [10^{10} M_{\odot}$ ]} & {$M_{\rm dyn} \sin^2 i \, [10^{10} M_{\odot}$]} & {$L_{\rm FIR} \, [10^{13} L_{\odot}$]} & {$L_{\rm bol} \, [10^{14} L_{\odot}$]} & {$\rm SFR \, [10^3 M_{\odot}/\rm yr$]} & {$M_{\rm dust}\,  [10^8 M_{\odot}$]} \\ 
\hline 
6.42 & $4.9 \pm 2.5 $ & $2.3 \pm 1.9 $ & $3.4 \pm 1.3$  & $2.2 \pm 0.33$ & $1.36 \pm 0.74$ & $2.0 \pm 0.5$ & $3.4^{+1.38}_{-1.54}$\\ 
\hline 
\end{tabular} 
}
\end{center}
\end{table*}

\section{The model}
\label{sec:model}

In this section we provide a brief summary of the original \textsc{GAMETE/QSOdust} model (referring the readers to \citealt{Valiante11, Valiante12, Valiante14} for a more detailed description) and we present the new features that have been implemented for the present study.

We reconstruct 30 independent merger histories of a dark matter halo at redshift 6.4 assumed to be the host of J1148. We adopt a Navarro Frenk $\&$ White (1995, NFW) density profile with a mass of $M_{\rm h} = 10^{13} M_{\odot}$, within the range supposed to host high-$z$ bright quasars \citep{Volonteri06,Fan04} and simulate its hierarchical history using a binary Monte Carlo merger tree algorithm based on the Extended Press-Schechter theory \citep{Lacey93}.\\
The code follows the time evolution of the mass of gas, stars, metals and dust in a 2-phase ISM inside each progenitor galaxy \citep[see also][]{deBennassuti14}, taking into account chemical enrichment from Asymptotic Giant Branch (AGB) stars and Supernovae (SNe), which inject dust and metals into the ISM, grain destruction by SN shocks and grain growth in dense molecular clouds. \\
Energy-driven outflows, powered by both AGN and SN feedback, are considered in the model: the energy released by the BH accretion process and SN explosions couples with the gas and can unbind a huge amount of interstellar gas \citep{Silk98}. Although the physical mechanisms that trigger these galaxy-scale winds are still controversial, the model predicts mass ejection rates comparable to the observed ones \citep{Maiolino12, Valiante12, Cicone15}.
 
Following \citet{Valiante11, Valiante16} we focus our study on one of the most distant and best studied quasar, J1148, discovered at redshift $z \simeq 6.4$ \citep{Fan03}. The observationally inferred properties of this
quasar are reported in Table \ref{Tab1}. These are used to calibrate the model by fixing the adjustable free parameters shown in Table \ref{Tab:free}, as
described below.

In what follows, we discuss the new features of the code, namely: ({\it a}) the formation of the disk via gas cooling;
({\it b}) the formation of the bulge via major mergers; ({\it c}) bursted and quiescent star formation both in the disk and in the bulge;
({\it d}) the BH seeding prescription; ({\it e}) the BH growth via accretion and coalescences, considering also the recoil velocities that can be generated by the product of the merging pair due to asymmetric gravitational wave emission; ({\it f}) SNe and AGN feedback, responsible of galactic-scale winds.

We adopt a Lambda cold dark matter ($\Lambda$CDM) cosmology with parameters $\Omega_m = 0.314$, $\Omega_\Lambda = 0.686$, $h=0.674 $ \citep{Planck14}, so that the Hubble time at redshift 6.4 is 851 Myr. The difference with the cosmological parameters adopted in previous works (Valiante et al. 2011, 2014) is mainly the larger value of $\sigma_8$ (Planck $\sigma_8 = 0.834 $, WMAP7 $\sigma_8 = 0.761$ ), which implies an increased power at small scales, leading to a larger number of progenitor systems at high redshifts.

\subsection{Gas cooling}

In each newly virialized dark matter halo with mass $M_h$, the initial gas mass is assumed to be the cosmic baryon fraction 
$M_{\rm \rm diff} = (\Omega_{\rm b}/\Omega_{\rm m}) \, M_{h}$.
We suppose this gas to be all in the diffuse phase, i.e. pressure-supported, and to follow an isothermal 
density profile $\rho_g$ defined as:

\begin{equation}
\rho_g(r) = \frac{M_{\rm diff}}{4\pi R_{\rm vir}r^2},
\end{equation}
\noindent
where $R_{\rm vir}$ is the virial radius of the dark matter halo.
The hot diffuse gas gradually cools providing the reservoir of cold gas out of which stars form. The gas cooling processes strongly depend on the temperature and
chemical composition of the gas. 

In dark matter halos with virial temperature $T_{\rm vir} < 10^4 \,K$,  referred to as mini-halos, the primordial gas can cool only through $\rm H_2$
roto-vibrational transitions \citep{Haiman96}. As the gas becomes progressively enriched in heavy elements, other
molecular species can contribute to cooling and collisionally excited metal fine-structure lines, mostly OI, CII can provide 
additional cooling pathways. 
Here we only consider the contribution of $\rm H_2$, OI and CII cooling using metallicity dependent tabulated cooling functions, $\Lambda(T_{\rm vir}, Z)$, 
computed as described in Appendix A of \citet{Valiante16} but we neglect the effect of $\rm H_2$ photo-dissociation by Lyman-Werner photons. We will
return to this point in Section \ref{sec:results}.  

In dark matter halos with virial temperatures $\rm T_{\rm vir} \geq 10^4 K$, referred to as Lyman-$\alpha$ cooling halos, the temperature is high enough to excite atomic transitions, allowing the primordial
gas to cool through hydrogen Lyman-$\alpha$ line emission. In this regime, we use metallicity-dependent tabulated cooling functions presented by
\citet{Sutherland93}.

The time scale for gas cooling, $\tau_{\rm cool}$, is defined as:

\begin{equation}
\tau_{\rm cool} = \frac{3}{2}\frac{ \mu m_p \kappa_B T_{\rm vir}}{\rho_g(r_{\rm cool}) \Lambda(T_{\rm vir},Z)},
\end{equation}

where $\kappa_B$ is the Boltzmann constant, $\mu$ is the mean molecular weight and $r_{\rm cool}$ is the cooling radius and it is obtained by assuming that the cooling time is 
equal to the halo dynamical time $t_{\rm dyn} = R_{\rm vir}/v_{\rm DM}$, where $v_{\rm DM}$ is the dark matter (DM) halo circular velocity:

\begin{equation}
r_{\rm cool} = \left[ \frac{t_{\rm dyn} \, M_{\rm diff} \, \Lambda(T_{\rm vir}, Z)}{6 \pi \, \mu m_p\, \kappa_B T_{\rm vir}  \,R^2_{\rm vir}}\right]^{1/2}.
\end{equation}
\noindent
Then, the gas cooling rate can be computed\footnote{Note that if $r_{\rm cool} > R_{\rm vir}$ we assume that the gas never reaches
hydrostatic equilibrium and it is immediately available to star formation \citep{Delucia10}.} as:

\begin{equation}
\dot{M}_{\rm cool} = 4 \pi \rho_g (r_{\rm cool}) r_{\rm cool}^2 \frac{dr_{\rm cool}}{dt} = \frac{M_{\rm diff}}{2R_{\rm vir}}\frac{r_{\rm cool}}{t_{\rm dyn}}.
\end{equation}

\subsection{Disk and bulge formation}
\label{diskBulge}

Along the hierarchical history of the final DM halo, we define major (minor) halo-halo merger events as those with halo mass ratio $\mu = M_{\rm halo,1}/M_{\rm halo,2}$ (with
$M_{\rm halo,1} \leq M_{\rm halo,2}$) larger (lower) than $\mu_{\rm thr} = 1/4$ \citep{Barausse12}. In quiescent evolution (i.e. no encounters with other galaxies), the cold gas settles on a rotationally-supported disk, because of the conservation of angular momentum, and can start to form stars. 
The disk, composed of gas and stars,
can be disrupted by a major merger and a spherical bulge is expected to form in this event.
Minor mergers, instead, are not expected
to destroy the disk but may help the growth of the bulge by disk instabilities \citep{Naab03,Bournaud05}.

In our model, major mergers are supposed to destroy both the gaseous and stellar disk components of the newly-formed galaxy, adding the stars and gas to the central bulge. Minor mergers do not contribute to the transfer of matter between the disk and bulge, and thus lead to the formation of a new galaxy with disk and bulge masses that are the sum of the two progenitors ones.

We consider a self-gravitating disk, with an exponential gas surface density profile, $\Sigma_{\rm d}$, defined as \citep{Mo98}:

\begin{equation}\label{sigma}
\Sigma_{\rm d}(r) = \Sigma_{\rm d}(0) \, e^{-r/R_{\rm d}} ,
\end{equation}
\noindent
where $R_{\rm d}$ is the scale radius of the gaseous disk and $\Sigma_{\rm d}(0)$ is the 
central surface densities of the gas. For the stellar component of the disk, we adopt the same
density profile with {the same scale radius} $R_{\rm d}$.
Following \citet{Mo98} we define the scale radius as,

\begin{equation}\label{rd}
R_{\rm d} = \frac{1}{\sqrt{2}}\left( \frac{j_{\rm d}}{m_{\rm d}} \right) \lambda R_{\rm vir} \frac{1}{\sqrt{f_{\rm c}}} f_{\rm R}(\lambda, c, m_{\rm d}, j_{\rm d}),
\end{equation}
\noindent
where $j_{\rm d} = J_{\rm d}/J$ is the ratio between the disk angular momentum and that of the halo, $m_{\rm d} = M_{\rm d}/M_{\rm h}$ is the disk mass (stars+gas) fraction over the halo mass. From the conservation of the specific angular momentum we assume $j_{\rm d} / m_{\rm d} = 1$. The spin parameter $\lambda$ is considered to be constant and equal to $0.05$, the mean value adopted by \citet{Mo98}.

The factors $f_{\rm c}$ and $f_{\rm R}$ take into account the correction to the total energy of the halo resulting from the NFW density profile and the gravitational effect of the disk, and are computed following the prescription given by \citet{Mo98}. The factor $f_{\rm c}$ depends on the concentration parameter $c$, that we assume to be constant and equal to $c=1$\footnote{Unfortunately,
numerical studies of the concentration parameter of dark matter halos spanning the mass and redshift range relevant for the present study are not available. Extrapolating the results of \citet{Munoz11},
we adopt a constant value of $c = 1$. At a fixed halo mass, BH growth would be favoured in more concentrated halos, that are characterized by a larger mass and circular velocity in the inner regions \citep{Mo98}.}:

\begin{equation}
f_{\rm c} = \frac{c}{2}\frac{1 - 1/(1+c)^2 - 2\ln(1+c)/(1+c)}{[c/(1+c) - \ln(1+c)]^2}.
\end{equation}

\noindent
The factor $f_{\rm R}$ is computed as,

\begin{equation}\label{fr}
f_{\rm R} = 2\left[ \int_0^{\infty} e^{-u}u^2\frac{v_c(R_{\rm d} u)}{v_c(R_{\rm vir})}\right]^{-1}, 
\end{equation}
\noindent
where $v_c(r)$ is the total rotation velocity of the system,

\begin{equation}
v_c^2(r) = v_d^2(r) + v_b^2(r) + v^2_{\rm DM}(r).
\label{eq:totcirc}
\end{equation}
\noindent
Here $v_b$ is the circular velocity of the bulge, $v_{\rm DM}$ is the circular velocity of the DM halo and $v_d$ is the circular velocity of the thin, exponential disk,

\begin{equation}
v^2_{d} = \pi \, G\,  \Sigma_0 \, x^2 [I_0(x/2)K_0(x/2) - I_1(x/2)K_1(x/2)],
\end{equation}
\noindent
where $x = r/R_{\rm d}$ and $I_{\alpha},K_{\alpha}$ are the modified Bessel functions of the first and second 
type, respectively and $\Sigma_0 = \Sigma(0)_{\rm d} + \Sigma(0)^{\star}_{\rm d}$ is the sum of the gas and stellar 
central ($r=0$) surface densities.

For the bulge component, we assume that the 
gas density profile $\rho_{\rm b}(r)$ is described as \citep{Hernquist90}, 

\begin{equation}\label{rho}
\rho_{\rm b}(r) = \frac{M_{\rm b}}{2\pi}\frac{r_{\rm b}}{r(r+r_{\rm b})^3},
\end{equation}
\noindent
where the scale radius, $r_{\rm b}$, is computed as $r_{\rm b} = R_{\rm eff}/1.8153$ (Hernquist 1990), and the effective radius $R_{\rm eff}$\footnote{$R_{\rm eff}$ is the effective radius of the isophote enclosing half the light.},  depends on the gas and stellar masses in the bulge \citep{Shen03}:
\begin{equation}\label{reff}
\log(R_{\rm eff}/{\rm kpc}) = 0.56\log(M_{\rm b} + M^{\star}_{\rm b}) - 5.54.
\end{equation}
\noindent
We adopt the same density profile for the stellar component in the bulge.

The velocity profile of the bulge, computed through the Poisson equation is

\begin{equation}
v^2_{b} = \frac{Gr(M_{\rm b} + M^{\star}_{\rm b})}{(r_{\rm b}+r)^2}.
\end{equation}
\noindent
We assume that the halo responds adiabatically to the gradual build up of the disk and bulge, maintaining the spherical symmetry during the contraction. Thus, the angular momentum is conserved during the collapse from a mean initial radius $r_i$ to a radius $r$ ($< r_i$), so that: 

\begin{equation}
M_f(r)r = M(r_i)r_i,
\end{equation}
\noindent
where $M(r_i)$ is the mass of DM enclosed in $r_i$ obtained 
integrating the NFW density profile and $M_f(r)$ is the total final mass within a radius r:

\begin{equation}
M_f(r) = M_{\rm d,t}(r) + M_{\rm b,t}(r) + (1-f_{\rm gal})M(r_i),
\end{equation}
\noindent
where $M_{\rm d,t}(r)$ and $M_{\rm b,t}(r)$ are the total disk and bulge masses (star and gas) enclosed within a
radius $r$, obtained by 
integrating eqs.~(\ref{sigma}) and~(\ref{rho}), and $f_{\rm gal} = [M_{\rm d,t} + M_{\rm b,t}]/M_{\rm h}$ is the
fraction of the total mass in the disk and bulge.

The velocity curve of the perturbed DM halo is then,
\begin{equation}
v^2_{\rm DM}(r) = [G(M_f(r)-M_{\rm d,t}(r)-M_{\rm b,t}(r)]/r.
\end{equation}
\noindent
Following these prescriptions we model the formation and evolution of disk and bulge components in each halo along the reconstructed merger histories.

\begin{table}
\begin{center}
\caption{The calibrated values of the adjustable parameters of the reference model.}
\begin{tabular}{|c|c|c|}
\hline 
\multicolumn{2}{|c|}{\textbf{Free parameters}} & \textbf{values}  \\ 
\hline 
$\epsilon_{\rm d}^{\star} $ & quiescent star formation efficiency  & $0.083$ \\ 
\hline 
$\beta$ & BH accretion efficiency & $0.03$ \\ 
\hline 
$\epsilon_{\rm AGN} $ & AGN-feedback efficiency & $1.5 \times 10^{-3}$ \\ 
\hline 
\end{tabular}\label{Tab:free}
\end{center}
\end{table}

\subsubsection{Star formation rate}

Hydrodynamical simulations suggest that merging events, major mergers in particular, can trigger 
bursts of star formation in the central regions as a consequence of the tidal forces produced by
galaxy-galaxy interactions \citep{Mihos94,Springel00,Cox08}.

Since starbursts are confined in the very central region of the galaxy, we assume a quiescent mode 
of star formation in the disk while bursts are triggered in the bulge when a major merger occurs.
The star formation rate (SFR) in the disk is described as,

\begin{equation}\label{SFR}
 \dot{M}^{\star}_{\rm d} =  \epsilon_{\rm d}^{\star} \frac{M_{\rm d},}{\tau_{\rm d}}
\end{equation}
\noindent
where $M_{\rm d}$ is the gas mass in the disk, $\tau_{\rm d}~=~3 R_{\rm d}/v_c(3R_{\rm d})$ is the dynamical time of the 
disk evaluated at the peak of the circular velocity profile \citep{Mo98}, $R_{\rm d}$ is the disk scale radius defined in eq. \ref{rd} and $\epsilon_{\rm d}^{\star}$ is an adjustable free parameter representing the star formation efficiency in the disk. In our reference model, $\epsilon_{\rm d}^{\star} =0.083$ (see Table \ref{Tab:free}).

Similarly, the SFR in the bulge is computed as,

\begin{equation}
\dot{M}^{\star}_{\rm b}= \epsilon_{\rm b}^{\star}  \frac{M_{\rm b}}{\tau_{\rm b}},
\end{equation}
\noindent
where $M_{\rm b}$ is the gas mass in the bulge, $\tau_{\rm b} = R_{\rm eff}/v_c(R_{\rm eff})$ is the dynamical time of the bulge and the effective radius $R_{\rm eff}$ is defined in eq. \ref{reff} above.
We assume that in absence of merger events, the star formation efficiency in the bulge is equal to that of the disk, 
$\epsilon_{\rm b}^{\star} =  \epsilon_{\rm d}^{\star}$. When a merger event occurs, the star formation efficiency increases as a consequence of the destabilizing effect of the interaction, and we adopt the following scaling relation:
\begin{equation}
\epsilon_{\rm b}^{\star} = \epsilon_{\rm d}^{\star}  \, f(\mu),
\label{eq:bulge_eff}
\end{equation}
\noindent
with $f(\mu) = \max[1,1+ 2.5 \, (\mu-0.1)]$, so that mergers with $\mu \le 0.1$ do not trigger starbursts. With the adopted scaling
relation, {the starburst efficiency in the reference model is $0.083 \le \epsilon_{\rm b}^{\star} \le 0.27$, consistent
with the range of values found by means of hydrodynamical simulations of merging galaxy pairs \citep{Cox08}
and adopted by other studies \citep{Menci04,Valiante11}.

\subsection{Black hole growth and feedback}
\label{Sec:bh growth}

\subsubsection{BH seeds}
We assume BH seeds to form only as remnants of first (Pop III) stars. In fact, our main aim is to investigate if SMBHs can form by super-Eddington accretion starting from ``light'' seeds at high redshift. Although the initial mass function of Pop III stars is still very uncertain, the most recent numerical simulations suggest a characteristic mass of a few hundreds of solar masses
at $z \sim 25$, that progressively shifts to a few tens of solar masses at lower redshifts \citep{Hirano15}. For simplicity, here we do not consider the redshift modulation of the characteristic mass
and we plant a BH seed with a mass of $M_{\rm seed} = 100 \, M_{\odot}$ in each newly-virialized halo with a metallicity $Z < Z_{\rm cr} = 10^{-4} Z_\odot$, above which the effects of dust and metal
line cooling allow the gas to fragment, reducing the characteristic mass to values comparable to those found in local stellar populations \citep{Schneider02,Schneider03,Schneider12a,Omukai05}.

\subsubsection{BH accretion}

Once formed, the BH accretes gas from the surrounding medium. The correlation between the mass of central SMBH and the bulge mass or velocity dispersion (\citealt{Magorrian98, Richstone98}, see \citealt{Kormendy13} and references therein) and the small scale on which the accretion takes place, suggest that the accretion onto the central black hole should be fuelled by the cold gas present in the bulge.  

The collapse of material onto the central BH in a galaxy is triggered by both merger-driven infall of cold gas, which loses angular momentum due to galaxy encounters, and quiescent accretion, assuming that the 
accretion rate is proportional to the cold gas mass in the bulge,

\begin{equation}
\dot{M}_{\rm accr} = \frac{f_{\rm accr} M_{\rm b}}{\tau_{\rm b}},
\end{equation}
\noindent
where, similarly to eq.~(\ref{eq:bulge_eff}), the accretion efficiency is expressed as,

\begin{equation}\label{accretion}
 f_{\rm accr} = \beta  \, f(\mu),
\end{equation}
\noindent
where $\beta$ is an adjustable free parameter. In our reference model, $\beta = 0.03$ (see Table \ref{Tab:free}),
{so that the efficiency of BH accretion is about $\sim 1/3$ of the efficiency of star formation in the bulge.}

Thus, the mass growth rate is,

\begin{equation}
\dot{M}_{\rm BH} = (1 - \epsilon_r) \dot{M}_{\rm accr},
\end{equation}
\noindent
where the radiative efficiency $\epsilon_r$ is defined as,

\begin{equation}
\epsilon_r = \frac{L_{\rm bol}}{\dot{M}_{\rm accr}\, c^2},
\end{equation}
with $L_{\rm bol}$ being the bolometric luminosity emitted by the accretion process.
At high accretion rates, the \citet{Shakura73} model of BH growth through a thin disk, where all
the heat generated by viscosity is immediately
radiated away, is incorrect. Instead, it is possible to use the optically thick, slim accretion disk solution, that is characterized by low radiative efficiencies \citep{Abramowicz88}. \\ 
The bolometric luminosity, $L_{\rm bol}$, is computed starting from the numerical solutions of the relativistic slim accretion disk equations obtained by \citet{Sadowski09}, adopting the fit presented by \citet{Madau14}: 
 
\begin{equation}
\frac{L_{\rm bol}}{L_{\rm Edd}} = \ A(a) \left[ \frac{0.985}{\dot{M}_{\rm Edd}/\dot{M}_{\rm accr} + B(a)} + \frac{0.015}{\dot{M}_{\rm Edd}/\dot{M}_{\rm accr} + C(a)} \right] ,
\label{eq:slimdisk}
\end{equation}
\noindent
where the Eddington accretion rate is defined as $ \dot{M}_{\rm Edd} \equiv 16 \, L_{\rm Edd} / c^2 $ and $A(a), B(a)$ and $C(a)$ are functions of the BH spin parameter $a$, 

\begin{eqnarray}
A(a) & = & (0.9663 - 0.9292 a)^{-0.5639} ,\\
B(a) & = & (4.627 - 4.445 a)^{-0.5524} ,\\
C(a) & = & (827.3 - 718.1 a)^{-0.7060}.
\end{eqnarray} 
\noindent
The slim accretion disk model represented by eq.~(\ref{eq:slimdisk}) predicts that even when the accretion rate is super-Eddington, with $1 \lesssim \dot{M}_{\rm accr}/\dot{M}_{\rm Edd} \lesssim 100$, the disk luminosity  remains only mildy super-Eddington, with $L_{\rm bol} \lesssim (2 - 4) \, L_{\rm Edd}$. In fact, in this regime a large fraction of the energy generated by viscosity does not have the time to be radiated away and is instead
advected into the black hole. As a result, the radiative efficiency is very small, with $0.002 \lesssim \epsilon_r \lesssim 0.05$, almost independently of the value of the BH spin parameter (see Figure 1 in \citealt{Madau14}.
Conversely, when the accretion rate is sub-Eddington, the radiative efficiency increases reaching an almost constant value which depends on the BH spin, as in the standard think disk solution, with $ \epsilon_r \lesssim 0.05$
for $a=0$ and $ \epsilon_r \lesssim 0.3$ for $a = 0.98$.

 Here we do not describe the time evolution of the BH spin parameter and we simply assume that the module of the spin vector $a$ is randomly extracted from a uniform distribution 
 \citep{Tanaka09, Barausse12}.

\subsubsection{BH mergers}

In halo merging events, we assume that the two nuclear BHs coalesce with the
same timescale of their host halos. However, in minor mergers (with $\mu < \mu_{\rm thr} = 1/4$, see Section \ref{diskBulge}) only the largest of the two progenitors BHs can settle in the centre of the new halo potential well, 
surviving as a nuclear BH, while the smaller one ends up as a satellite.

During the BH merger, the newly formed BH receives a large center-of-mass recoil due to the net linear momentum carried by the asymmetric gravitational waves emission \citep{Campanelli07, Schnittman08, Baker08}. 
The recoil (or kick) velocity of the coalesced binary depends on the mass ratio of the merging pair and on the amplitude and orientation of the spin vectors of the two BHs. Here we follow the parametrization presented 
by \citet{Tanaka09} and - for each merger event - we compute the kick velocity as a function of the BH mass ratio assuming the spin vectors to be randomly oriented. The average kick velocities increase
with the mass ratio of the merging pair, $q = M_{\rm BH,1}/M_{\rm BH, 2}$ (with $M_{\rm BH,1} \leq M_{\rm BH, 2}$). For strongly unequal mass mergers, with $0.01 \lesssim q \lesssim 0.1$, we find $\left\langle v_{\rm kick} \right\rangle = 1 - 100$ km/s, whereas
for larger mass ratios, with  $0.1 \lesssim q \lesssim 1$, the kicks can be very strong, with velocities  $\left\langle v_{\rm kick} \right\rangle = 100 - 1000$ km/s.

We then compare the kick velocity with the circular velocity at the radius of influence of the BH, $R_{\rm BH} = GM_{\rm BH}/v_c^2(R_{\rm BH})$ with $v_c(r)$ given by eq.~(\ref{eq:totcirc}),
and we retain the BH only when $v_{\rm kick} < v_c(R_{\rm BH})$. For $M_{\rm BH}/M_{\rm h} = 10^{-3}$, the retention velocity is $v_c(R_{\rm BH}) \sim 2 v_{\rm vir}$, where $v_{\rm vir}$ is the escape velocity at the virial radius  \citep{Yoo04}.
%



\subsubsection{BH feedback}

There is now strong observational evidence that the energy released by the quasar can drive powerful galaxy-scale outflows
(for recent works see \citealt{Feruglio15, Carniani15, Cresci15} and references therein). 
Outflowing gas at velocities up to $v \sim 1400$ km/s traced by [CII] emission has been detected in SDSS J1148
(Maiolino et al. 2012) with an estimated total mass outflow rate of $1400 \pm 300 \, M_\odot/$yr 
that decreases with distance from the quasar,
ranging from a peak value of $\sim 500 \, M_\odot/$yr at $\sim 3$~kpc to $\lesssim 100 \, M_\odot/$yr at $\sim 20$~kpc 
\citep{Cicone15}.  

In \citet{Valiante12} we show that the quasar-driven mass outflow rate predicted by \textsc{GAMETE/QSOdust},
on the basis of a simple energy-driven wind, is in good agreement with the observations. 
Here we follow a similar approach,  adopting the so-called “blast wave” model, 
in which the AGN radiation field can accelerate the gas generating fast supersonic winds which propagates 
outwards through an expanding blast wave, pushing out the surrounding medium 
(see e.g. \citealt{Cavaliere02, King03, King05, King10, Lapi05, Menci05, Menci08,
Zubovas12, Zubovas14, Costa14} and references therein).

In this framework, the energy released by the AGN that couples with the interstellar gas is estimated as,

\begin{equation}\label{feedback}
\dot{E}_{\rm AGN} = \epsilon_{\rm AGN} \, \epsilon_r \, \dot{M}_{\rm accr} c^2,
\end{equation}
\noindent
where the coupling efficiency $\rm \epsilon_{AGN}$ is an adjustable free parameter. In our reference model $\rm \epsilon_{AGN} = 1.5 \times 10^{-3}$ (see Table \ref{Tab:free}). 

If the post shock material does not cool efficiently, 
the bubble expands adiabatically and the outflow is energy-driven.
As the blast wave propagates from the center of the halo, it first interacts with the gas of the disk and bulge, 
reheating a fraction of cold gas and transferring mass to the diffuse hot phase.

When the shock has propagated beyond the bulge and disk radius, part of the gas mass is ejected from the galaxy, if the binding energy is not enough to hold the material.


The mass outflow rate at a given radius $r$ can be estimated as: 
\begin{equation}
\dot{M}_{\rm w, AGN} (r) =  2 \, \epsilon_{\rm AGN} \, \epsilon_r \, \left(\frac{c}{v_c(r)}\right)^2 \dot{M}_{\rm accr},
\label{eq:outflow}
\end{equation} 
\noindent
where $v_c$ is the circular velocity of the system given by eq.~(\ref{eq:totcirc}), and we evaluate the above
equation at the bulge, disk and DM halo virial radius. 

A similar description is used to describe the effects of SN-driven winds. 
The mass outflow rate beyond a given radius $r$ is given by:
\begin{equation}
\dot{M}_{\rm w, SN} (r) =  \frac{2 \, \epsilon_{\rm SN}  \, E_{\rm SN}}{v_c(r)^2} \, R_{\rm SN}
\end{equation} 
\noindent
where $R_{\rm SN}$ is the rate of SN explosions, $E_{\rm SN}$ is the average SN explosion energy, and $\epsilon_{\rm SN} = 1.6 \times 10^{-3}$
is the SN wind efficiency \citep{Valiante12}. The time-dependent SN rate and explosion energy is computed for each galaxy along the merger tree according to
formation rate, age and initial mass function of its stellar population. A detailed description of the chemical evolution model can be found in
\citet{Valiante11, Valiante14} and \citet{deBennassuti14}.

\begin{figure}
\centering
\includegraphics[width=8cm]{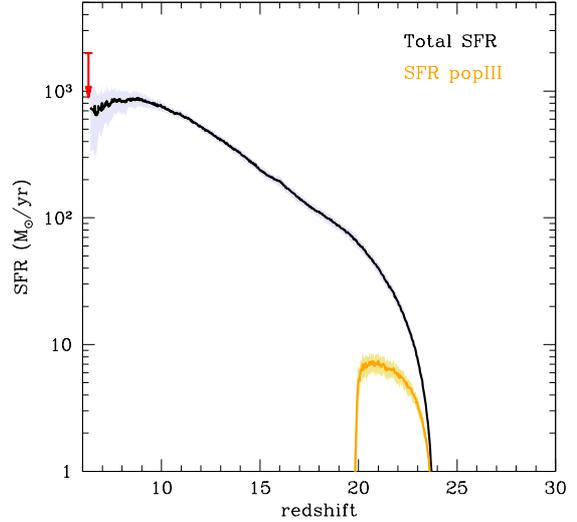}
\caption{Redshift evolution of the total SFR (black line) and of Pop III stars (orange line), averaged over the 29 realizations. Shaded areas represent 1-$\sigma$ dispersions and the red arrow indicates the upper limit on the SFR inferred from the IR luminosity (see in the text for further details).}
\label{figure:sfr}
\end{figure}

\section{Results}
\label{sec:results}

In this section, we present the predicted evolution of the hierarchical assembly of the SMBH and its host galaxy. To explore the dependence of 
the results on the population of progenitors and their merger rate, for the same model parameters we have run 30 independent merger trees. 
In one merger tree we find that a merger occurs at $z = 6.43$ between two black holes of $M_{\rm 1, BH} = 1.7 \times 10^9 M_{\odot}$ and $M_{\rm 2, BH} = 1.6 \times 10^9 M_{\odot}$, 
producing a recoil velocity $\sim 2$ times higher than the retention speed, $v_c(R_{\rm BH})$. The newly formed BH is displaced from the center and it stops accreting gas. For this
reason, we do not consider this to be a viable formation route for a bright quasar like J1148, and we exclude this merger tree from the sample average.

\subsection{The formation of stars and BH seeds}

In Fig.~\ref{figure:sfr},  we show the redshift evolution of the total SFR (summed over all the progenitor galaxies in each simulation) and the separate contribution of Pop~III stars.
We also show the upper limit on the SFR of $\sim 2000 \, M_{\odot}/$yr (Table \ref{Tab1}) inferred from the observed FIR luminosity using the relation
$L_{\rm FIR}/L_{\odot} = 10.84 \times 10^9 {\rm \, SFR}/(M_{\odot}/$yr) \citep{Valiante14}. This relation\footnote{The conversion factor between the FIR luminosity and the SFR has been 
obtained assuming a 10 - 200 Myr burst of stars with solar metallicity and a Larson IMF with $m_{\rm ch} = 0.35 M_\odot$
\citep{Valiante14}.} is based on the assumption of 
starburst dominated dust heating and it provides only an upper limit to the real SFR, due to the non-negligible contribution from the AGN. According to a recent detailed radiative transfer analysis,
the AGN can provide up to 60\% of the total FIR luminosity \citep{Schneider15}, decreasing the SFR by a factor 1.4 - 2.5, in agreement with the average value of $\sim 800 \, M_\odot$/yr predicted
by the reference model. 

Due to efficient metal enrichment, Pop~III star formation becomes negligible below $\rm z \sim 20$ and no more BH seeds are formed, consistent with other studies (\citealt{Madau01,Haiman01,Heger03,Volonteri03,Madau04, Valiante16}. The mass distribution of DM halos which host BH seeds ranges between 
$\sim 3 \times 10^6 M_{\odot}$ and $\sim 10^8 M_{\odot}$ with a peak at $M_{\rm h} \sim 10^7 M_{\odot}$, as shown in Fig.~\ref{seed_m}. 
Thus, we find that a major fraction ($\sim 90\%$, on average) of BH seeds are formed in DM mini-halos,
where gas cooling could be easily suppressed due to H$_2$ photo-dissociation by Lyman-Werner photons. 
The inclusion of this additional feedback effect slows down metal enrichment and extends BH seeds formation to lower
redshifts ($z \geq 15$) and larger DM halos ($\sim 5 \times 10^7 - 10^9 M_{\odot}$). While the evolution of the total BH
mass and BH accretion rate at $z < 15$ is only mildly affected, the birth environment of late-forming seed BHs
(gas rich Ly-$\alpha$ cooling halos) may be more favorable to SE accretion.  Here we do not consider the effect of H$_2$ photo-dissociation,
which we defer to a future study, and we assume that the formation rate of Pop~III stars is limited only by metal enrichment. 
\begin{figure}
\centering
\includegraphics[width=8cm]{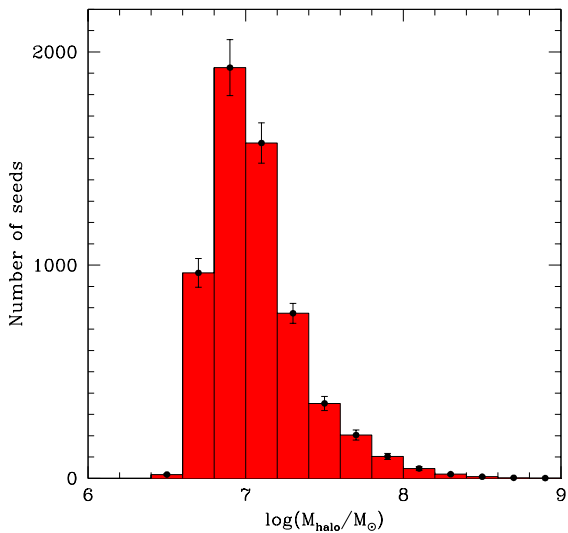}
\caption{Mass distribution of halos hosting a newly formed 100 $M_{\odot}$ BH seed, averaged over the 29 realizations with 1-$\sigma$ error bars.}
\label{seed_m}
\end{figure}  

\subsection{BH evolution}

\begin{figure*}
\includegraphics[width=8.4cm]{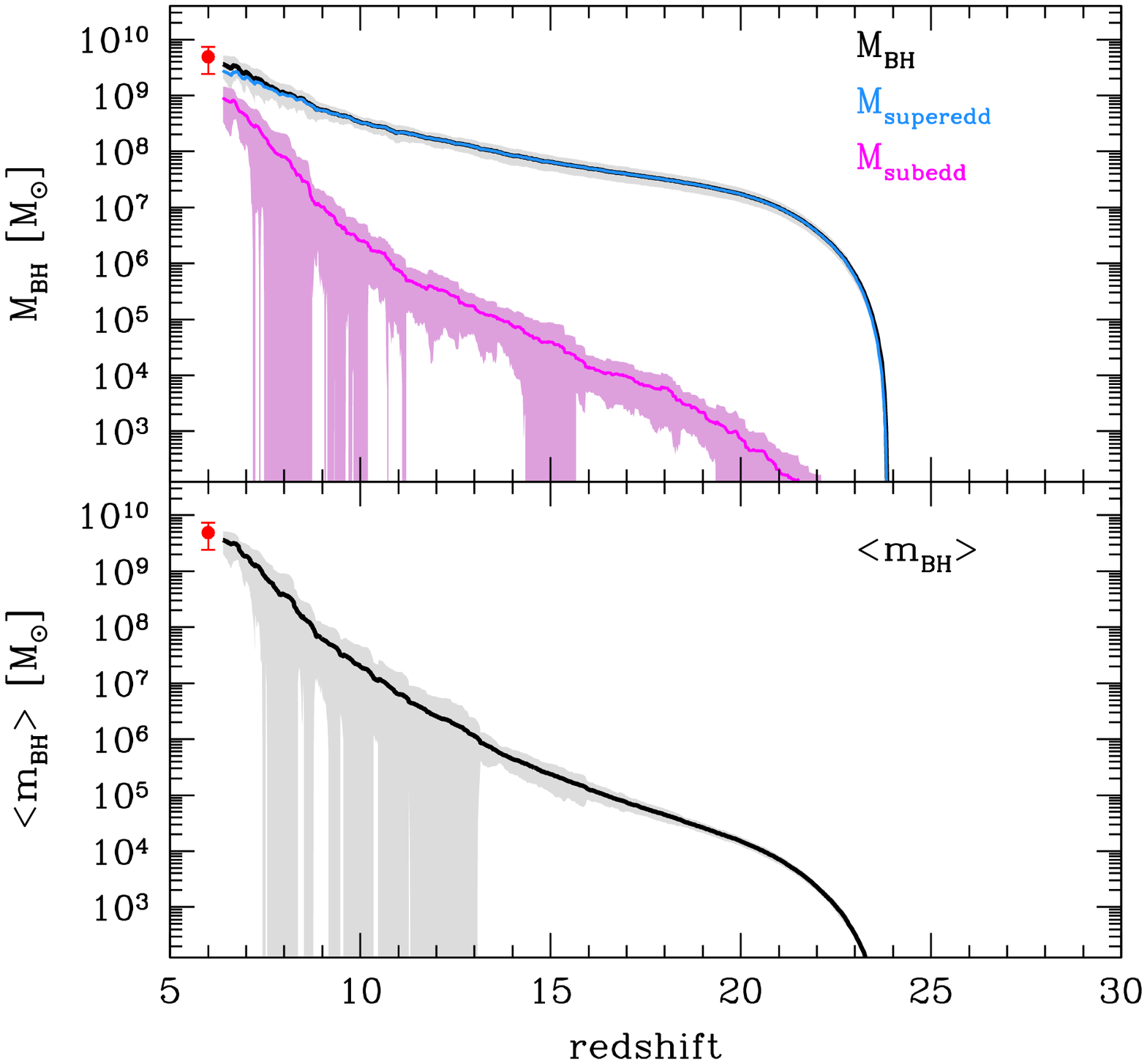}
\includegraphics[width=8.4cm]{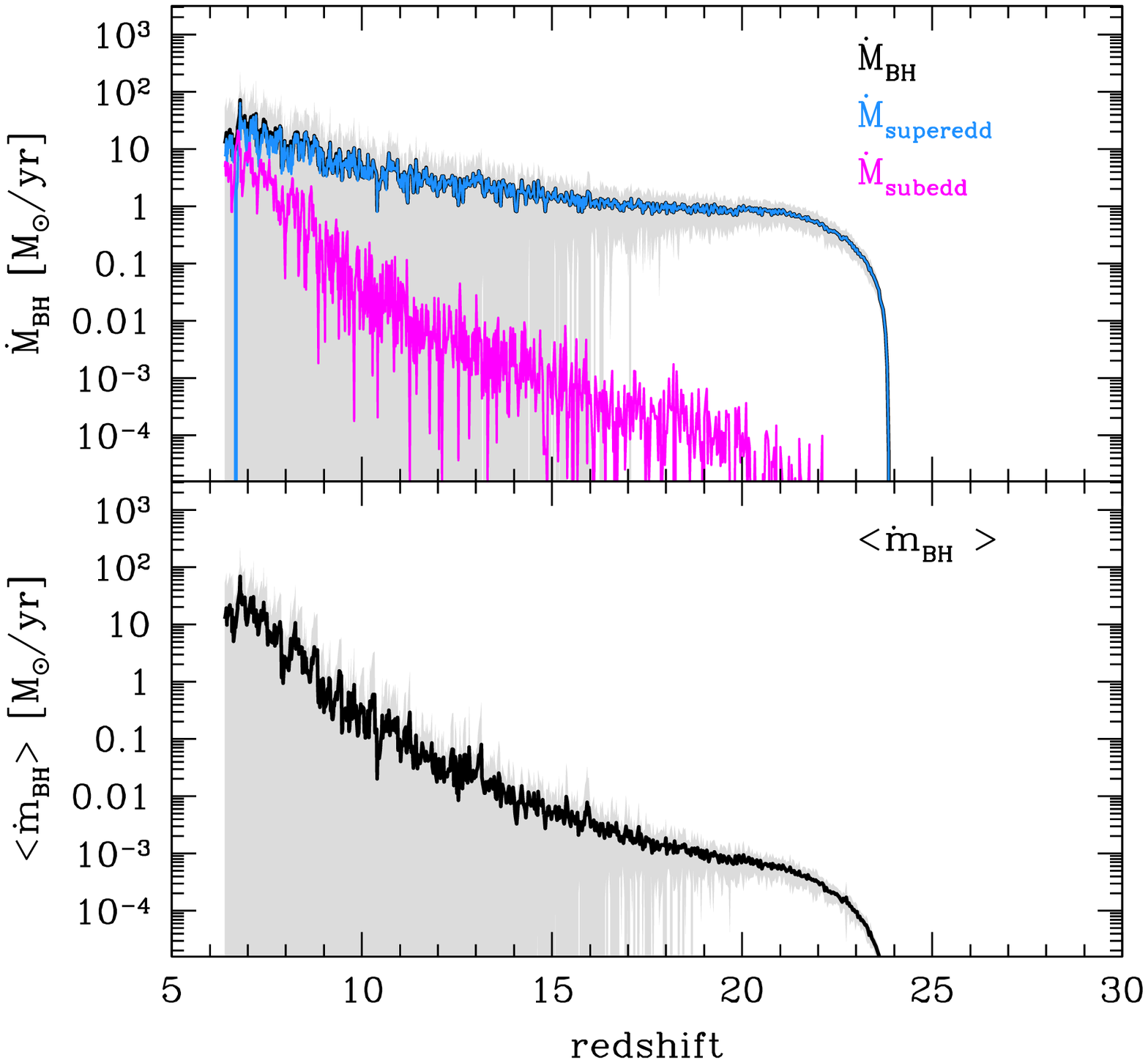}
\caption{\small{Redshift evolution of the total and mean BH masses and BHARs, averaged over 29 independent  merger trees. Shaded areas are 1-$\sigma$ dispersions.
\textit{Top, left panel}}:  total BH mass (summed over all 
BH progenitors at each redshift in each simulation, black line) and the BH mass grown by means of sub-Eddington (magenta line) and super-Eddington (cyan line) accretion events. 
\textit{Top, right panel}:  total BHAR (black line) and BHAR obtained considering only sub- (magenta line) and super- (cyan line) Eddington accreting BHs.
The mean BH mass and BHAR (averaged over all BH progenitors at each redshift in each simulation) are shown in the bottom panels (left and right, respectively). }
\label{fig:BHevo}
\end{figure*}

In Fig.~\ref{fig:BHevo} we show the redshift evolution of the BH mass and black hole accretion rate (BHAR) predicted by our reference model. In the top panels, the values are obtained summing over all BH
progenitors present at each redshift in each simulation and then averaged over the 29 realizations. The different lines allow to separate the contribution to the BH mass and accretion rate achieved by means of sub-Eddington
($\le 16 \, L_{\rm Edd} / c^2$) and super-Eddington ($> 16 \, L_{\rm Edd} / c^2$) 
accretion events. By construction, the final BH mass predicted by the reference model is $\sim (3.6 \pm 1.6)\times 10^9 M_\odot$, in agreement with the value inferred from observations
of J1148 (see Table 1). We find that, on average, $\sim 75\%$ of the final SMBH mass grows
by means of super-Eddington gas accretion. This provides the dominant contribution to the total BHAR at all but the smallest redshifts. Although the quantities shown in all panels have been
averaged over 29 merger trees, the redshift evolution of the BHAR appears to be very intermittent, a consequence of rapid depletion/replenishment of the bulge gas reservoir out of which the
BHs accrete. 

To gain a better idea of the typical values of BH mass and BHAR predicted by the reference model,  in the bottom panels of Fig.~\ref{fig:BHevo} we also show the mean
quantities, averaged over all BH progenitors present at each redshift in each simulation. It is clear that at $20 \lesssim z \lesssim 25$ the mean BH mass rapidly grows from $\sim 100 \, M_\odot$
to $\sim 10^4 \, M_\odot$ by means of super-Eddington gas accretion rates of $10^{-5} M_\odot/{\rm yr} \lesssim {\rm BHAR} \lesssim 10^{-3} M_\odot/{\rm yr}$. Hence, due to early efficient 
super-Eddington accretion, the mean BH progenitors at $z \sim 20$ have already achieved a mass comparable to the BH mass predicted by the direct collapse scenario. 
This is consistent with what recently found by \citet{Lupi15} by means of high-resolution numerical simulations, which show that stellar-mass black holes can increase their
mass by 3 orders of magnitudes within a few million years while accreting gas at super-Eddington rates in the dense cores of high-$z$ galaxies. 

\begin{figure*}
\includegraphics[width=8.4cm]{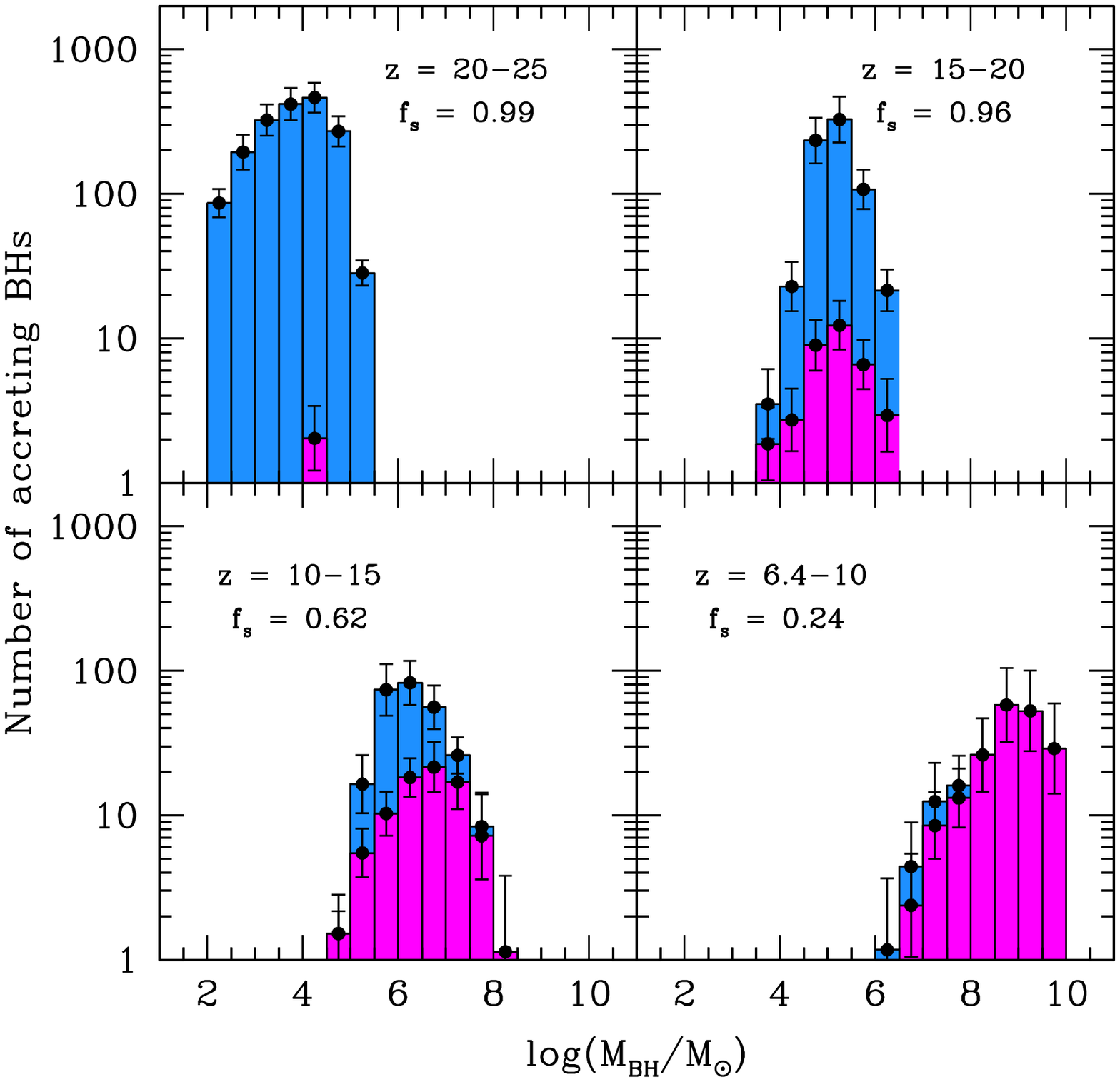}
\includegraphics[width=8.4cm]{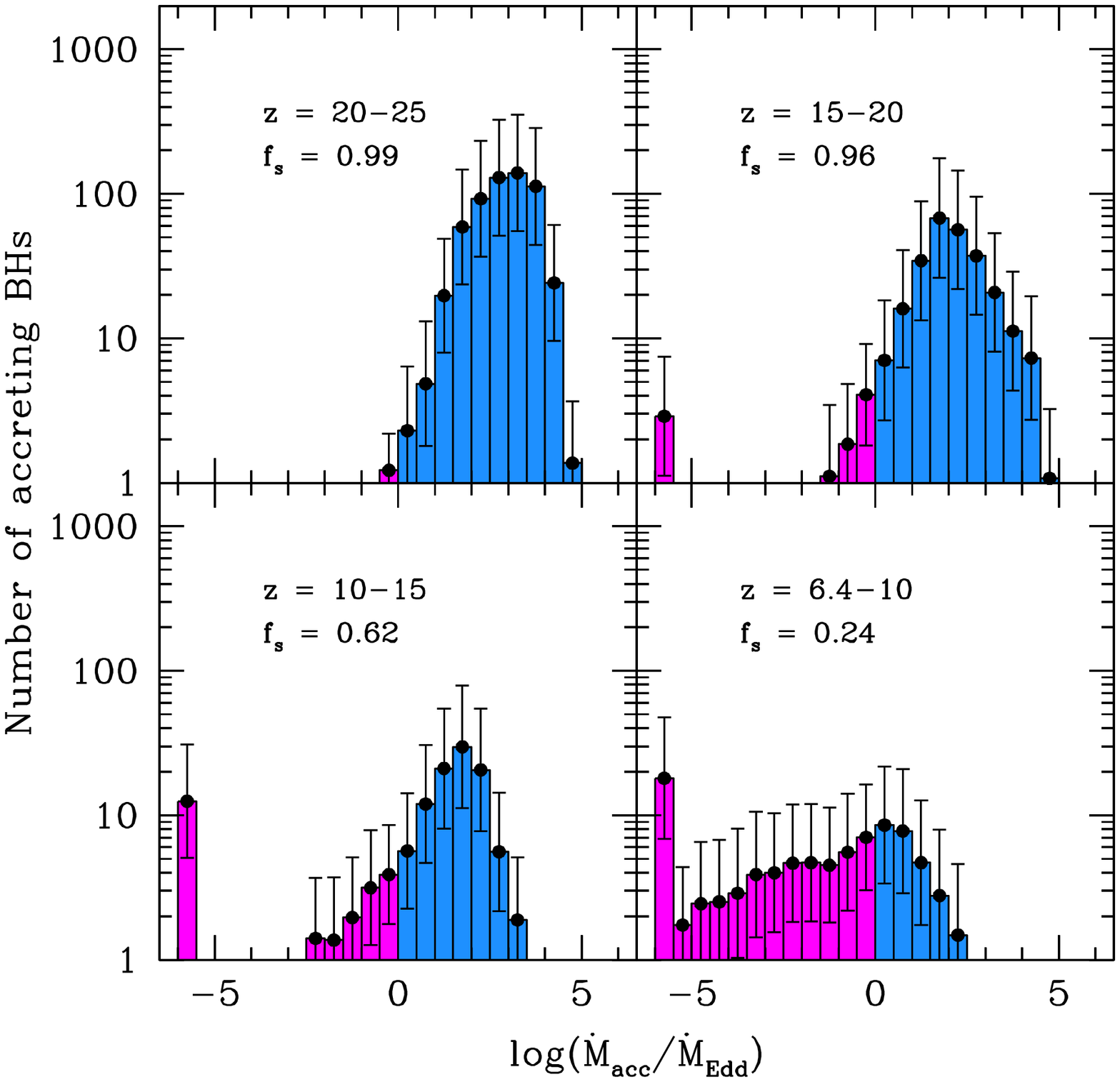}
\caption{\small{Number of accreting BHs as a function of the black hole mass ({\it left panel}) and the accretion ratio ({\it right panel}), averaged over 29 realizations with 1$-\sigma$ error bars.
The histograms show the number of super- (cyan) and  sub- (magenta) Eddington accreting BHs. In each figure, we separately show 4 different redshift intervals 
and we give the corresponding number fraction of super-Eddington accreting BHs over the total, $f_{\rm s}$.}}
\label{fig:histo}
\end{figure*}
\begin{figure}
\includegraphics[width=8cm]{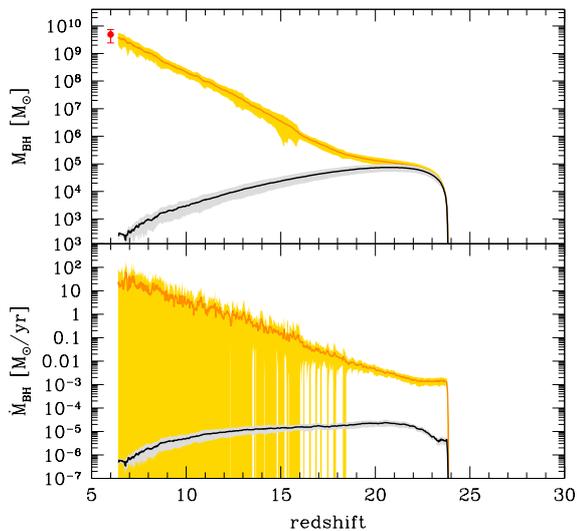}
\caption{\small{Redshift evolution of the total BH mass ({\it upper panel}) and BHAR ({\it lower panel}), averaged over 29 independent  merger trees. Shaded areas are 1-$\sigma$ dispersions.
In each panel, the orange line indicates the predicted evolution assuming $\dot{M}_{\rm accr} \leq 20 \, \dot{M}_{\rm Edd} = 320 \, L_{\rm  Edd}/c^2$ and the black line shows the evolution
assuming the conventional Eddington limited accretion, $\dot{M}_{\rm accr} \leq  L_{\rm  Edd}/c^2$ (see text).}}
\label{fig:BHevoEdd}
\end{figure}
\begin{figure}
\includegraphics[width=8cm]{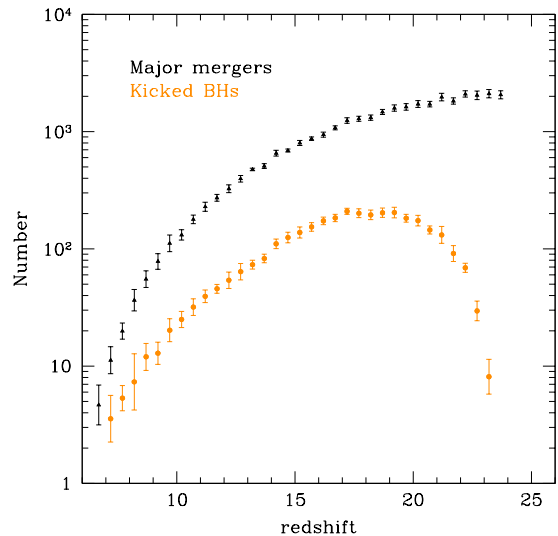}
\caption{\small{The average redshift distribution of major mergers (black triangles) and of kicked BHs during BH-BH coalescences in the
model where $\dot{M}_{\rm accr} \leq  L_{\rm  Edd}/c^2$ (orange points). Each point has been obtained averaging over 29 different merger tree realizations and the 
errorbars correspond to the 1-$\sigma$ dispersion.}}
\label{fig:BHkicks}
\end{figure}
\begin{figure*}
\centering
\begin{minipage}{.45\textwidth}
\includegraphics[width=\textwidth]{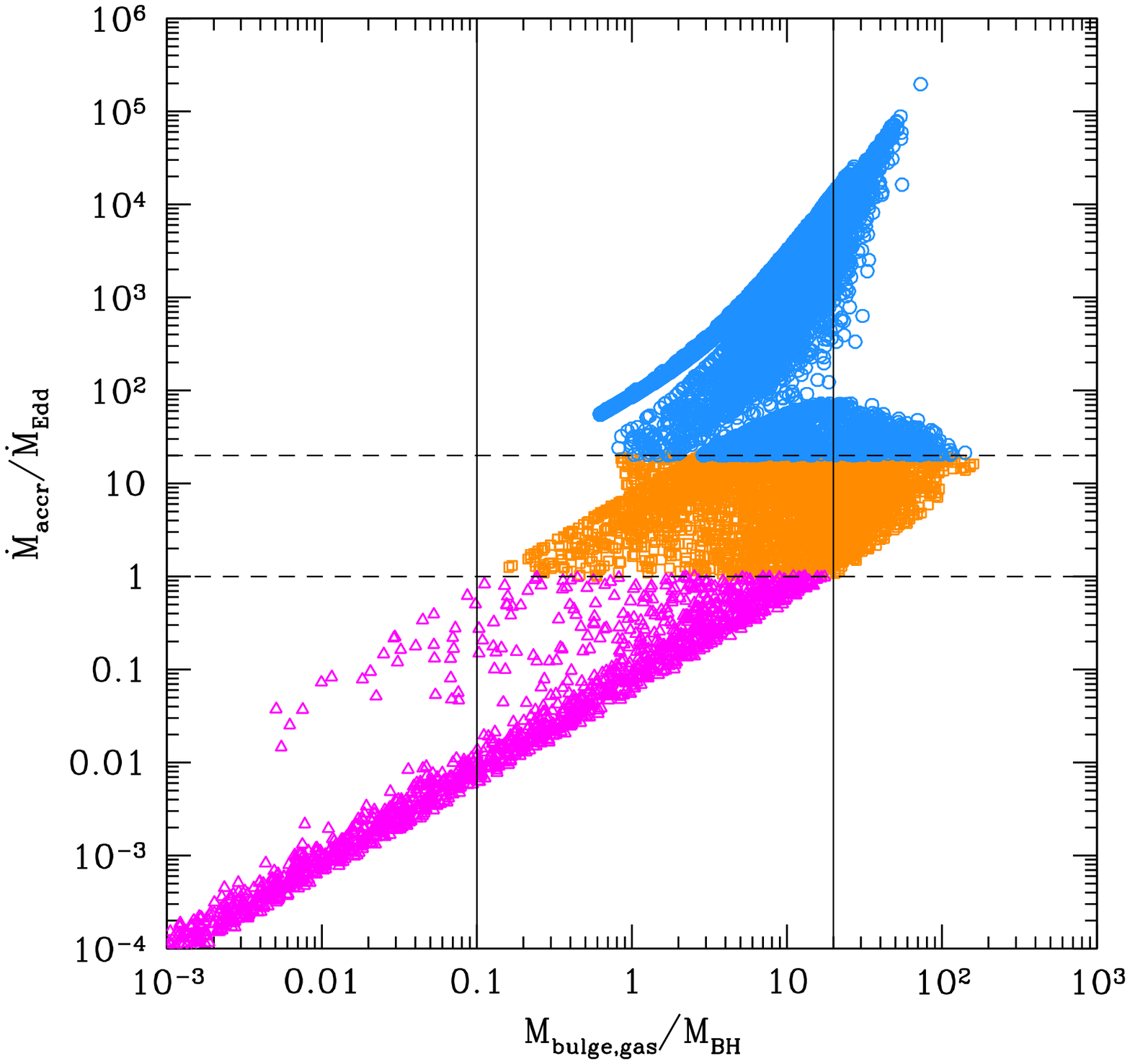}
\end{minipage}
\hspace{0mm}
\begin{minipage}{.45\textwidth}
\includegraphics[width=\textwidth]{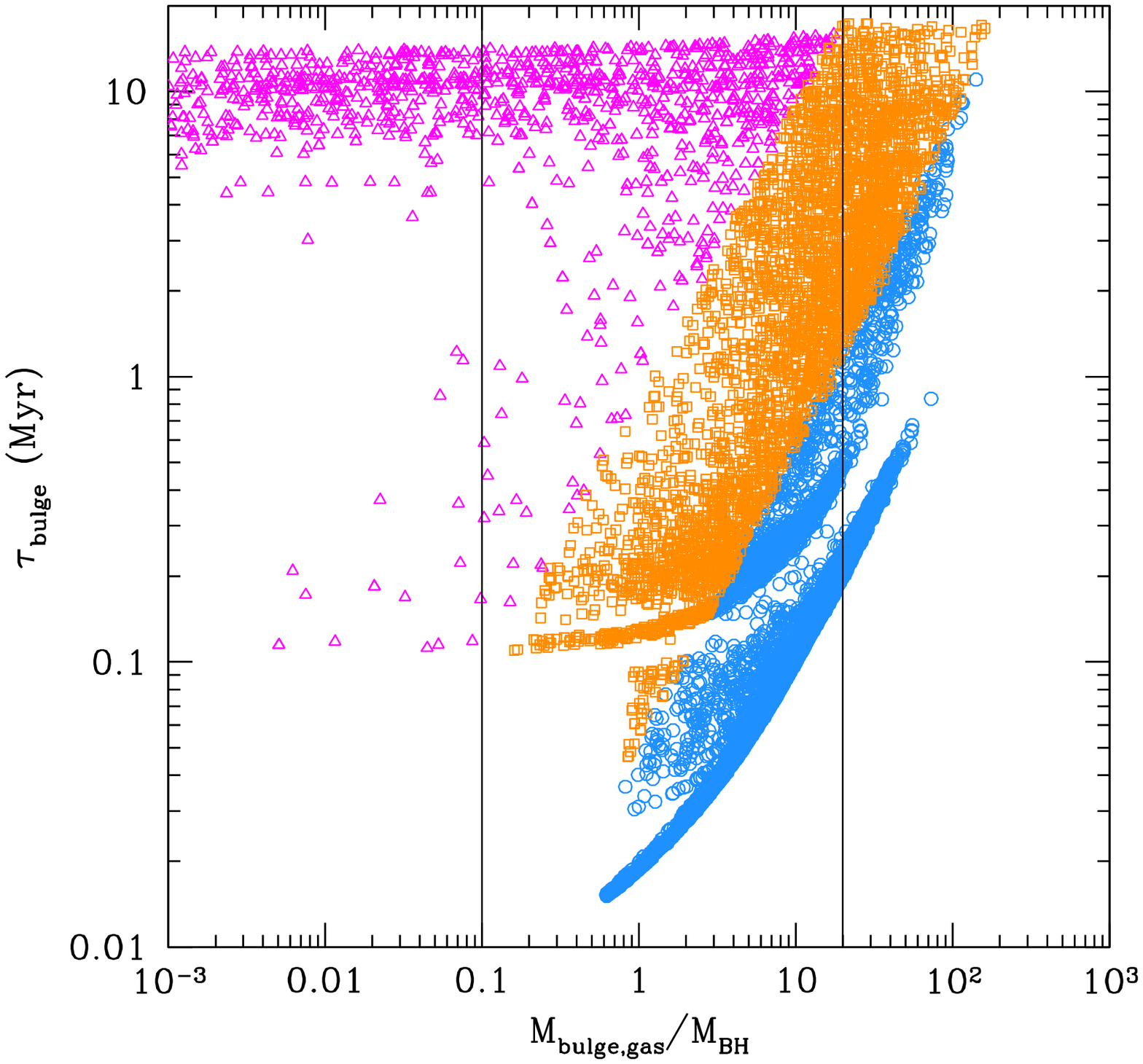}
\end{minipage}
\caption{\small{Eddington accretion ratio, $\dot{M}_{\rm accr}/\dot{M}_{\rm Edd}$, (\textit{left panel}) and dynamical timescale of the bulge, $\tau_{\rm b}$, (\textit{right panel}) as a 
function of the bulge gas - BH mass ratio, $M_{\rm b}/M_{\rm BH}$. Each point represents an accreting BH in any of the 29 merger histories. Sub-Eddington accreting
BHs are shown by magenta triangles, and we separate mildly super-Eddington accreting BHs with $1 \leq \dot{M}_{\rm accr}/\dot{M}_{\rm Edd} \leq 20$ (orange squares) and hyper-Eddington
accreting BHs with $\dot{M}_{\rm accr}/\dot{M}_{\rm Edd} > 20$ (cyan circles). The two horizontal dashed lines in the left panel allow to visually separate these regimes.
The vertical lines in both panels give two reference values of $M_{\rm b}/M_{\rm BH} =0.1$ and $20$ (see text).}}
\label{fig:prop}
\end{figure*}

Fig.~\ref{fig:histo} shows the average distribution of BHs accreting at super- and sub-Eddington rates as a function of the BH mass and Eddington accretion ratio for different redshift intervals. 
The reference model predicts that, at $15 \le z \le 25$, almost all BH progenitors accrete at super-Eddington rates. Since the BH masses are still
relatively small, $10^2 \, M_\odot \le M_{\rm BH} \le 10^6 \, M_\odot$, BH accretion rates of $10^{-5} M_\odot/{\rm yr} \lesssim {\rm BHAR} \lesssim 5 \times 10^{-3} M_\odot/{\rm yr}$,
which characterize the early mass growth (see the bottom right panel of Fig.~\ref{fig:BHevo}), correspond to very large accretion ratios, $\dot{M}_{\rm accr}/\dot{M}_{\rm Edd} \sim 10^2 - 10^4$. 
The mass of BH progenitors increases with time and the fractional number of super-Eddington accreting BHs decreases, being $f_{\rm s} \sim 60\%$ at $z \sim 10-15$ and 
dropping to $f_{\rm s} \sim 20\%$ at $z < 10$. Because of the larger BH masses, the accretion ratios are smaller and $\dot{M}_{\rm accr}/\dot{M}_{\rm Edd} <500$ at $ z < 10$.

For most of the evolution, we find that BH progenitors accrete at highly super-Eddington rates, with $\dot{M}_{\rm accr}/\dot{M}_{\rm Edd} >> 10$.
At these large Eddington accretion ratios the
applicability of the adopted slim disk solution is highly debated. In fact, recent general-relativistic magneto-hydrodynamical simulations show that BHs accreting at  
$20 < \dot{M}_{\rm accr}/\dot{M}_{\rm Edd} < 200$ develop a disk structure that is still radiatively inefficient, with total luminosities that do not exceed $\sim 10 \, L_{\rm Edd}$, but
the total energy escaping the system can be very large, mostly in the form of thermal and kinetic energy of outflowing gas and Poyinting flux \citep{McKinney14,Sadowski13}.
However, \citet{Inayoshi15} have shown that there exist regimes where steady accretion rates larger than 3000 times the Eddington rate can be sustained. 

To better assess the impact of these extreme hyper-Eddington accretion events on our results, we have run the same set of simulations discussed so far but artificially imposing an
upper limit of $\dot{M}_{\rm accr} \leq 20 \, \dot{M}_{\rm Edd} = 320 \, L_{\rm  Edd}/c^2$ to the gas accretion rate. The results are shown in Fig.~\ref{fig:BHevoEdd}. In the same figure, we also show, for comparison, the evolution
predicted  assuming Eddington-limited accretion. In order to better compare with previous results, this model has been run 
assuming $\dot{M}_{\rm accr} \leq L_{\rm Edd}/c^2$ ($1/16$ smaller than the definition adopted in the present study, see
Eq.~\ref{eq:slimdisk}), as conventionally adopted in the literature. 

We find that, even when the Eddington accretion ratio is  $\dot{M}_{\rm accr}/\dot{M}_{\rm Edd} \leq 20$, the 
final SMBH mass predicted by the reference model is in good agreement with the observations. The high-redshift evolution of both the total BH mass and the total BHAR, however,
is markedly different from the results shown in Fig.~\ref{fig:BHevo}. At $z > 10$ the BHAR is several orders of magnitudes smaller and the BH mass is correspondingly affected,
being $\sim 10^6 \, M_\odot$ at $z \sim 15$ ($\sim 1/100$ of the total BH mass shown in Fig.~\ref{fig:BHevo} at the same $z$). Due to the smaller gas accretion rates at high redshifts, a larger gas fraction is retained around nuclear BHs at $z < 10$. As a result,
the BH mass has a steeper late growth rate, with short episodes of intense gas accretion reaching $\sim 10^2 \, M_\odot/{\rm yr}$ at $z \sim 7$.

On the contrary, when Eddington-limited gas accretion is assumed, the final BH mass can no longer be reproduced using the reference model. In this case, the gas accretion rates are
too small to trigger fast BH growth at high redshifts. The total BH mass is dominated by the coalescence of BH seeds and its redshift evolution is strongly affected by lack of BH
seeds at $z < 20$ (see the behaviour of the Pop~III SFR in Fig.~\ref{figure:sfr}) and by kicks received during BH-BH coalescences in major mergers. 
Fig.~\ref{fig:BHkicks} shows the evolution of the average number of major mergers and of kicked BHs predicted by
the model. While the average number of major mergers decreases with time, the number of kicked BHs increases at $20 \lesssim z \lesssim 25$ and than decreases at lower $z$.
This is due to the combination of the growing number of BH seeds formed at high $z$  and of
the shallow potetial wells of their host mini-halos, which allow the kick velocity of the newly formed BH to easily exceed the retention speed. 

Hence, we can conclude that super-Eddington accretion is fundamental for the formation of the first SMBHs at $z > 6$,  even when extreme hyper-Eddington accretion
events are not considered.

\subsection{Environmental conditions for Super-Eddington accretion}

Our model enables us to perform a statistical study of the physical properties of the environments where BH progenitors accrete at super-Eddington rates. 
The left panel of Fig.~\ref{fig:histo} shows that when both sub- and super-Eddington accreting BHs are present, their BH masses are
comparable, with a tendency of sub-Eddington accreting BHs to have larger masses at lower $z$. Similarly, the occurrence of super-Eddington
accretion is not correlated with the mass of the host dark matter halo, nor with its gas content or metallicity. At each given value of any of 
these quantities, in fact, both sub- and super-Eddington accreting BHs are found in the simulations.  

The different accretion regimes are more cleanly separated when we plot the Eddington gas accretion ratio as a function of the ratio between the
gaseous bulge and the BH masses (see the left panel of Fig.~\ref{fig:prop}). Most of the BHs that accrete at sub-Eddington rates
are characterized by $M_{\rm b}/M_{\rm BH} < 20$, whereas the number of super-Eddington accreting BHs is negligible when
$M_{\rm b}/M_{\rm BH} < 0.1$. However, when $0.1 \leq M_{\rm b}/M_{\rm BH} \leq 20$ (the region
between the two vertical lines in the plot), the BHs can be characterized by vastly different accretion ratios: a good fraction
of the hyper-Eddington accreting BHs are found in this region of the plot. The larger
accretion rate in these systems is due to the much shorter dynamical time of the bulge. This is shown in the right panel of Fig.~\ref{fig:prop}.
A sequence of increasing bulge dynamical times is evident, with most of the BHs found in bulges with $0.01 \, {\rm Myr} \lesssim \tau_{\rm b} < 1 \, {\rm Myr}$ in hyper-Eddington,
$0.1 \, {\rm Myr} \lesssim \tau_{\rm b} < 20 \, {\rm Myr}$ in mildly super-Eddington, and $5 \, {\rm Myr} \lesssim \tau_{\rm b} < 20 \, {\rm Myr}$  in
sub-Eddington accretion regimes. Indeed, hyper-Eddington accreting BHs are predominantly found in high-$z$ systems, with less massive and more compact
bulges. The figure also shows that super-Eddington accretion requires gas-rich bulges and that, when $M_{\rm b}/M_{\rm BH} < 0.1$, only
sub-Eddington accreting BHs in massive, gas poor bulges are found.

The environmental conditions for super-Eddington accretion that emerge from our statistical study are in good agreement with the results recently
found by \citet{Lupi15}. By means of detailed hydro-dynamical simulations, these authors show that, in order to accrete at super-Eddington
rates, BHs must be embedded in dense gas structures, with masses comparable or larger than the masses of the accreting BHs.

\subsection{BH-driven outflow}

Outflowing cold gas in J1148, traced by [C {\small II}] emission, was first detected  by \citet{Maiolino12} with the 
IRAM Plateau de Bure Interferometer, and then confirmed with high-resolution follow-up observations by \citet{Cicone15}.
The outflow has a complex morphology and spatial extent, reaching a maximum projected
radius of 30~kpc. The estimated mass outflow rate and velocity are shown in Fig.~\ref{fig:outflow} as a function of the projected distance from the nucleus.  
In the same figure, we also show the predictions of the reference model. Following eq.~(\ref{eq:outflow}), 
the outflow velocity is computed as the circular
velocity at the corresponding radius, $v_{\rm w, AGN}(r) = v_{\rm c}(r)$, and
we estimate the mass outflow rate accounting for the delay
$\tau_{\rm dyn} = r/v_{\rm w, AGN}$ between the BH energy release and the observation. 
Due to the large variability of the BH luminosity, the 1-$\sigma$ dispersion among the
different merger trees of the predicted average mass outflow rate (gray shaded region in the upper panel)
is consistent with the data. However, the average values (black solid line) are larger than
observed and show a different radial dependence, especially at $r > 20$\,kpc. 
The bottom panel shows that the observed outflow travels at a velocity 
consistent with the  circular velocity of the host system. There are a few radii 
where the observed values are larger, probably reflecting a stronger coupling between the 
energy and momentum injected by the AGN and the surrounding gas. Yet, even if we take the
observed values of outflow velocities at each radius to estimate $\tau_{\rm dyn}$ and $\dot{M}_{\rm w, AGN}$
(see the blue dashed line in the upper panel with the cyan shaded region), the resulting 
mean mass outflow rate is still larger than observed. Our description of an energy-driven
wind with constant coupling efficiency may not be adequate to capture the complex dynamics of this
massive outflow. However, \citet{Cicone15} stress that the data should be considered as
a lower limit on the total mass outflow rate, because it accounts only for the atomic gas phase of
the outflow, while a significant amount of the outflowing mass may be in the molecular phase. 
\begin{figure}
\centering
\includegraphics[width=0.4\textwidth]{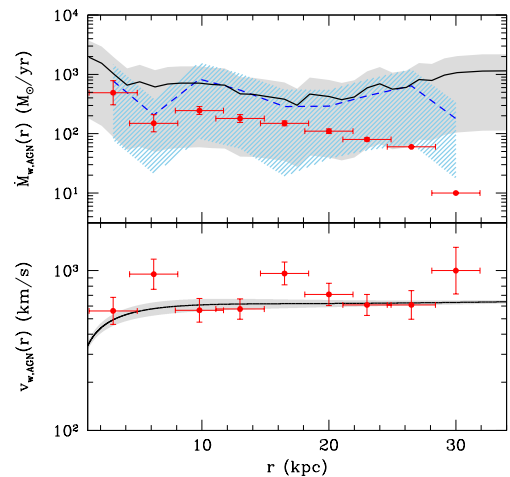}
\caption{The mass outflow rate ({\it upper panel}) and velocity ({\it lower panel}) as a function of the projected distance from the
nucleus.  \citet{Cicone15} observations are shown with red data points and the predictions of the reference model are shown by
black solid lines with shaded gray
regions. The blue dashed line in the upper panel (with the cyan dashed region) shows the predicted outflow rate that we would infer
using the BH luminosity predicted by the reference model and the observed outflow velocities (see text). The lines show the average
among the 29 merger trees and the shaded regions are the 1-$\sigma$ dispersion.}
\label{fig:outflow}
\end{figure}

\subsection{The coevolution of BHs and their host galaxies}

It is interesting to explore the implications of our results for the co-evolution of nuclear BHs and their host galaxies. In Fig.~\ref{fig:mrel} we show the evolutionary path
(from the bottom left to the top right) in the mean BH mass - stellar bulge mass ($\langle m_{\rm BH}\rangle$ - $\langle m_{\rm b}^{\star}\rangle$) plane predicted by the reference model (black solid line) and by the model with 
$\dot{M}_{\rm accr} \leq 20 \, \dot{M}_{\rm Edd}$ (orange solid line). In each simulation, we consider the mean values among all the SMBH progenitors
and their hosts present at each redshift, and then we average over the 29 merger trees. 
For comparison, we also show in the same figure the observational data and the empirical fit (gray data points and dashed line) for local galaxies provided by \citet{Sani11},
and the more recent scaling relation inferred for local ellipticals and classical bulges by \citet[][solid green line and shaded region]{Kormendy13}.

In the reference model, BH progenitors of the first SMBHs at $z > 6$ follow a symbiotic evolution, with a small offset with respect to the observed local scaling relation.
When $\dot{M}_{\rm accr} \leq 20 \, \dot{M}_{\rm Edd}$, the different evolution at high-$z$ is reflected in a steeper relation between the mean BH mass and the stellar
bulge, very close to that predicted by \citet{Kormendy13}. The difference between the models becomes negligible when $\langle m_{\rm BH}\rangle \, >10^7 \, M_\odot$ ($\langle m_{\rm b}^{\star}\rangle \, > 10^9 \, M_\odot$), which occurs - on average - at $ z \sim 10$.
 
When the average BH mass has reached its value of $(3.6 \pm 1.6) \times 10^9 M_{\odot}$ at $z = 6.4$, the host galaxy has already grown to a bulge (total) stellar mass of 
$2.7 \, (3.2) \times 10^{11} M_{\odot}$. Hence, we predict a final average BH-to-bulge (total) stellar mass ratio of $M_{\rm BH}/M_{\rm star} = 0.013 
\, (0.011)$, well within the scatter of the relations inferred from various observational studies of massive local galaxies \citep[][and references therein]{Marconi03, Sani11, Kormendy13}. However, this ratio is $\sim 25$ times smaller than what is inferred from observations of J1148
(red data point). Following the procedure commonly applied to high-$z$ bright QSOs, the stellar mass is computed as $M_{\rm star} = M_{\rm dyn} - M_{\rm H_2}$, with $M_{\rm dyn}$ and 
$M_{\rm H_2}$ inferred from CO observations (see Table 1, \citealt{Walter04,Wang10}). Similar results obtained for a larger sample of $z > 6$ QSOs
have suggested the idea that the first SMBHs grow faster than their host galaxies (\citealt{Wang10,Wang13,Venemans15} see however \citealt{Willott15}).

As suggested by \citet{Valiante14},  observations of high-$z$ QSOs are sensitive to the innermost $2.5 - 3$~kpc and may be missing a significant fraction of the galaxy
\citep{Valiante14}. This is also supported by recent observations of J1148, which show extended [C {\small II}] 158 $\mu$m emission and far-infrared (FIR) continuum,
likely associated with cold gas and star formation on scales of $\sim 10 - 20$~kpc \citep{Cicone15}.

Indeed, the mean bulge effective radius at $z = 6.4$ predicted by the model is $R_{\rm eff} = 7.3 \pm 0.8$~kpc, in
good agreement with observations of local galaxies hosting the largest BHs (see Fig.~\ref{fig:reff}). When we restrict to the innermost 2.5 kpc, we 
find a mean bulge stellar mass of $(3.9 \pm 0.2)\times 10^{10} M_\odot$, much closer to the observation (see the arrow and black data point in Fig.~\ref{fig:mrel}).
The same is true if we consider the mean gas mass within 2.5 kpc, that we predict to be $M_{\rm H_2} = (2.0 \pm 0.9) \times 10^{10} \, M_\odot$, that well reproduce
the observed value (see Table 1).

Finally, the reference model predicts a mean dust mass at $z = 6.4$ of $M_{\rm dust} = (3.6 \pm 0.9)\times 10^8\, M_\odot$, in good agreement with the value inferred
 from the FIR luminosity. This result has been obtained using the chemical evolution module, which includes dust processing in a 2-phase ISM, that 
 has been developed by \citet{Valiante11,Valiante14} and \citet{deBennassuti14}. Hence, consistent with previous findings \citep{Valiante11,Valiante14}, we
 find that the large dust mass that has enriched the ISM of the host galaxy is the result of a large stellar component, and that the apparent tension with the 
 observed dynamical mass - the so-called {\it stellar mass crisis} - is at least partly due to the small spatial extent of the observations. We refer the interested
 readers to \citet{Valiante14} for an extended discussion on this point.

\begin{figure}
\centering
\includegraphics[width=0.4\textwidth]{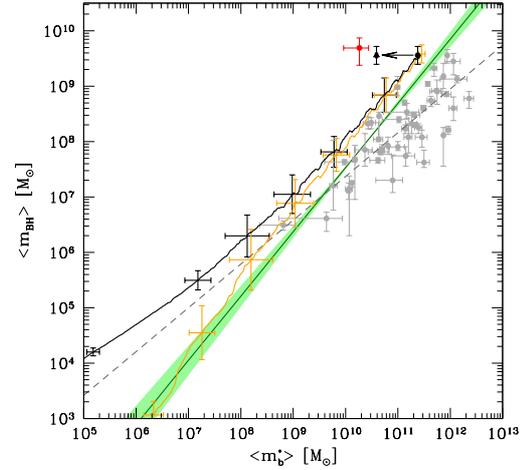}
\caption{Redshift evolution of the mean black hole mass as a function of the mean bulge stellar mass in SMBH progenitors for the reference model (black solid line) and
the model with $\dot{M}_{\rm accr} \leq 20 \, \dot{M}_{\rm Edd}$ (orange solid line). Gray circles are data for local galaxies, with the empirical fit (gray dashed line) provided by 
\citet{Sani11}. The solid green line with shaded region is the scaling relation derived by \citet{Kormendy13}. The red point represents the black hole and stellar mass within a region of 2.5 kpc inferred from observations of J1148 (Table \ref{Tab1}). 
The model predictions are averaged over 29 merger tree realizations and the errorbars show the 1-$\sigma$ dispersion for both mean BH and bulge stellar mass, at few selected redshift along the averaged merger histories. 
The arrow illustrates the reduction in stellar mass if we restrict to the central 2.5 kpc region (black data point, see text).}
\label{fig:mrel}
\end{figure}

\begin{figure}
\centering
\includegraphics[width=0.4\textwidth]{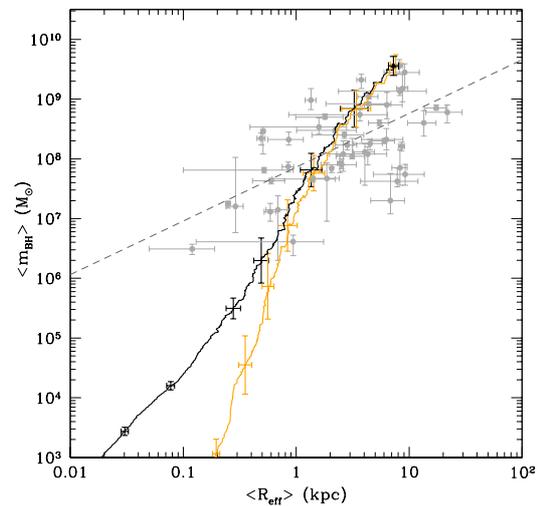}
\caption{Redshift evolution of the mean black hole mass as a function of the mean bulge effective radius of the host galaxy, averaged over 29 merger tree realizations with 1-$\sigma$ errorbars at few selected redshift,
for the reference model (black solid line), and the model  with $\dot{M}_{\rm accr} \leq 20 \, \dot{M}_{\rm Edd}$ (orange solid line). 
Gray circles represent data for local galaxies, with the empirical fit (gray dashed line) given by \citet{Sani11}.}
\label{fig:reff}
\end{figure}

\section{Discussion and conclusions}
\label{sec:discussion}

The data-constrained model GAMETE/QSOdust allows us to explore a large number of formation histories of a given quasar, in the present case  J1148 at $z = 6.4$,
reproducing the observations of the quasar and its host galaxy. With the adjustable free parameters that we have selected, described in Table 2, the model reproduces
the physical quantities listed in Table 1. 
 Hence, the properties that we predict for the host galaxy of J1148 (SFR, dust mass, gas and stellar masses) are consistent with previous results
obtained by (\citealt{Valiante14, Valiante16}) for the same quasar. 

With respect to \citep{Valiante11,Valiante14, Valiante16}, the current version of \textsc{GAMETE/QSOdust} enables to ({\it i}) follow the formation and
evolution of the disk and bulge in each progenitor galaxy, and ({\it ii}) remove the constraint of Eddington-limited BH accretion. 

In particular,  \citet{Valiante16} find that the formation of a few (between 3 and 30 in the
reference model) heavy BH seeds with masses $M_{\rm BH} = 10^5 \, M_\odot$ enables the Eddington-limited growth of a SMBH by $z = 6.4$. 
This conclusion heavily depends on the occurrence - among the progenitors - of Lyman-$\alpha$ cooling halos where gas cooling is suppressed 
by the low-metallicity and strong Lyman-Werner background \citep{Valiante16}. This ``head start'' requires favourable conditions, that are easily
erased by the joint interplay of chemical, radiative and mechanical feedback effects. 

Here we have explored the alternative scenario where the BHs can grow through a radiatively inefficient slim disk at super-Eddington rates.
This condition is easily met by light BH seeds formed in gas-rich systems at high redshifts.  

In the model presented in this work, we plant light BH seeds in newly virialized halos above redshift $z \sim 20$, before the effects of chemical feedback inhibit the formation of metal poor ($Z<Z_{\rm cr}$) stars. 
With this seeding prescription, we find that:

\begin{itemize}
\item On average, $\sim 80\%$ of the SMBH mass of J1148 is provided by super-Eddington gas accretion ($>16 \, L_{\rm Edd} / c^2$).
This represents the dominant contribution to BH growth down to $z \sim$ 10;

\item Due to fast and efficient super-critical accretion, the mean BH mass at redshift $z \sim 20$ is $\gtrsim 10^4 \, M_\odot$, comparable that predicted for heavy BH seeds formed by direct collapse; 

\item More than $90\%$ of BH progenitors accrete at super-Eddington rates at $15<z<25$ in dense, gas-rich environments. At these redshifts, hyper-Eddington accretion events, with $\dot{M}_{\rm accr}/\dot{M}_{\rm Edd} \sim 10^2-10^4$, are common;

\item The observed SMBH mass of J1148 at $z = 6.4$ can be reproduced even adopting a maximum super-Eddington accretion rate of $\dot{M}_{\rm accr} \leq 20 \, \dot{M}_{\rm Edd}$, showing that hyper-critical accretion
is not required;
 
\item BH progenitors of the final SMBH evolve in symbiosis with their host galaxies. The predicted AGN-driven mass outflow rate at $z = 6.4$ shows a radial profile that is broadly consistent with the lower limits
inferred from CII observations by Cicone et al. (2015); 

\item The predicted final BH-to-bulge (total) stellar mass ratio, 
$M_{\rm BH}/M_{\rm star} = 0.013 \, (0.011)$, is within the scatter of the observed local relation
and a factor of $\sim 25$ lower than inferred from dynamical mass observations of J1148.
The discrepancy is significantly reduced if we account
for the mass within 2.5\,kpc from the nucleus, the region targeted by CO data. At this radius,
the mean bulge stellar mass is $(3.9 \pm 0.2) \times 10^{10} \, M_\odot$, much closer to the 
observational value.
\end{itemize}

As a consequence of the lower gas accretion rates, 
the average BH mass predicted by \citet{Valiante16} is much smaller than in our reference model, at all but the latest redshifts
(see their Fig.~3). 
This difference is reduced when we impose that $\dot{M}_{\rm accr} \leq 20 \, \dot{M}_{\rm Edd}$. 
In this case, the average BH progenitor mass at $z \sim 15$ is comparable in the two models. However, while in \citet{Valiante16} the mass
growth is triggered by the formation of heavy seeds, in our model this is achieved by mildly super-Eddington accretion on
light BH seeds. 

The progenitors of SMBHs at $z > 6$ experience the {\it strong} form of coevolution
defined by \citet{Kormendy13}, where galaxies affect BH growth by controlling BH feeding and merging, 
and BHs control galaxy properties via AGN feedback.
In fact, while the small radiative efficiencies of super-Eddington accreting BHs 
is indispensable to limit the effects of AGN feedback \citep{Lupi15}, at $z > 10$ 
the BHs shine at a few Eddington luminosities with a noticeable effect 
on the cold gas content of their host galaxies. At lower $z$, an increasing fraction of BH 
progenitors accrete at sub-Eddington rates, but with larger radiative efficiencies. As a result of the larger BH
mass and BH accretion rates, AGN-driven winds at $z < 10$ power strong galaxy-scale outflows and 
suppress star formation, leading to the down-turn of the total SFR shown in Fig.~\ref{figure:sfr}.

\begin{figure}
\centering
\includegraphics[width=0.4\textwidth]{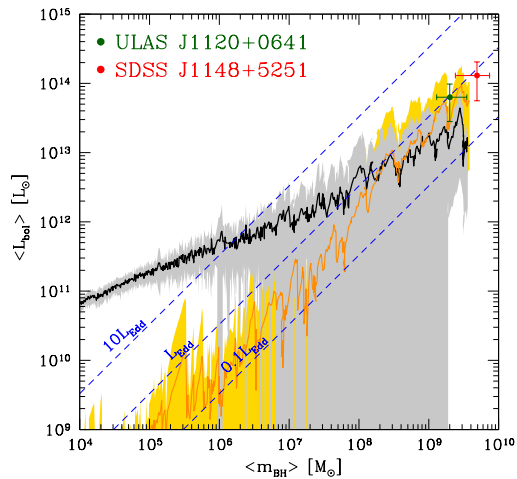}
\caption{Mean bolometric luminosity of BH progenitors as a function of the mean BH mass predicted by the
reference model (black solid line) and by the model with $\dot{M}_{\rm accr} \leq 20 \, \dot{M}_{\rm Edd}$ (yellow solid line). 
For each model, the lines show the average among the 29 merger trees and the shaded regions are the 1-$\sigma$ dispersion.
The data points show the observational values of the two quasars SDSS J1149 (red circle) and ULAS J1120 (green square). 
The diagonal dashed lines show some reference values of the luminosity in units of the Eddington luminosity.}
\label{fig:Lbol}
\end{figure}

In Fig.~\ref{fig:Lbol} we show the average bolometric luminosity as a function of the average BH mass 
of SMBH progenitors for the reference model (black solid line) and for the model
with $\dot{M}_{\rm accr} \leq 20 \, \dot{M}_{\rm Edd}$ (yellow solid line).  The model predictions are compared 
with observations of SDSS J1148 ($z = 6.4$) and of the most distant quasar
currently known, ULAS J1120 at $z=7.1$ \citep{Mortlock11}. The errorbars on the bolometric
luminosities account for the observational uncertainties on the flux at $1450 \, {\AA}$ and on the bolometric
corrections \citep{Richards06}.
Some reference values of the luminosity in units of the Eddington luminosity are shown by the
diagonal dashed lines. The difference among the two models reflects the different BH accretion
history: in the model with $\dot{M}_{\rm accr} \leq 20 \, \dot{M}_{\rm Edd}$ the first BH progenitors
accrete at a lower rate, saving cold gas for the latest evolutionary phases. As a result, 
for BH progenitors with $M_{\rm BH} \lesssim 10^8 \, M_\odot$, the mean luminosity 
predicted by the reference model is always super-Eddington (with $L_{\rm bol} > 10\, L_{\rm Edd}$
when $M_{\rm BH} \lesssim 10^6 \, M_\odot$), whereas in the model with  $\dot{M}_{\rm accr} \leq 20 \, \dot{M}_{\rm Edd}$
the mean luminosity is always $0.1 \, L_{\rm Edd} < L_{\rm bol} < L_{\rm Edd}$. However, in the latest
evolutionary phases, when $M_{\rm BH} > 10^8 \, M_\odot$, this trend is reversed. 
Given the observational uncertainties and the large variability among different merger trees,  the luminosity of J1148 
is consistent with the model predictions. Interestingly, the data point of ULAS J1120 is also lying within the 1-$\sigma$
dispersion. Indeed, we find that $\sim 20\%$ of BH progenitors at $z = 7.1$ have luminosities and masses compatible
with the observed values of ULAS J1120, indicating that this quasar may be one of the progenitors of SDSS J1148 at $z = 6.4$.

\section*{Acknowledgments}
We thank Valeria Ferrari, Nicola Menci and Marta Volonteri for useful discussions and comments.
The research leading to these results has received funding from the European Research Council under the European 
Union’s Seventh Framework Programme (FP/2007-2013)~/~ERC Grant Agreement n. 306476.

\bibliography{Bib}
\label{lastpage}

\end{document}